Masayuki UEHARA
Department of Physics, Saga University,
Saga 840,  Japan
\documentstyle[12pt]{article}
\renewcommand{\theequation}%
     {\thesection.\arabic{equation}}
\begin{document}
\baselineskip 24pt
\begin{flushright}
SAGA-HE-96\\
March 1996\\
\end{flushright}
\par

\begin{center}
{\LARGE\bf Meson-hyperon couplings \\
{\hfil \hfil}\\
in the bound-state approach \\
{\hfil \hfil}\\
to the Skyrme  model}
\end{center}
\vspace{.5cm}
\begin{center}
{\large Y. KONDO}\\
{\sl Suzuka University of Medical Science and Technology, 
Suzuka 510-02}\\
{\large S. SAITO\footnote{E-mail address 
saito@nuc-th.phys.nagoya-u.ac.jp} 
and Y. TANUSHI\footnote{E-mail address 
tanushi@nuc-th.phys.nagoya-u.ac.jp}}\\
{\sl Department of Physics, Nagoya University, Nagoya 464-01}\\
{\large M. UEHARA\footnote{E-mail address ueharam@cc. saga-u.ac.jp}}\\
{\sl Department of Physics, Saga University, Saga 840}
\end{center}
\vspace{0.5cm}
\noindent{\bf Abstract}\par
Kaon and pion coupling constants to hyperons are calculated  in the 
bound-state approach to strangeness in the 
Skyrme-soliton model. The pion and kaon coupling constants  
are properly defined as matrix elements of 
source terms of the mesons sandwiched between two single-baryon 
states.  Numerical calculation of the coupling constants shows that 
the bound-state approach well reproduce the empirical values. 
\newpage
%%%%%%%%%%%%%%%%%%%%%%%%%%%%%%%%%%%%%%%%%%%%%%%%%%%%%%%%%%%%%%%%%%%%%%%%
% Definitions
\def\CH{{\cal H}} \def\CJ{{\cal J}} \def\CK{{\cal K}} 
\def\CL{{\cal L}}\def\CM{{\cal M}} \def\CV{{\cal V}} 
\def\CD{{\cal D}}
\newcommand\beq{\begin{equation}}
\newcommand\eeq{\end{equation}}
\newcommand\beqa{\begin{eqnarray}}
\newcommand\eeqa{\end{eqnarray}}
\newcommand\noeq{\nonumber}
\def\dspl#1{\displaystyle{#1}}
\def\noi{\noindent}
\def\Ref#1{\,[{#1}]}
\def\Tr{{\rm Tr}}
\def\Nc#1{O($N_c^{#1}$)}
\def\SU#1{$SU({#1})$}
\def\finv{\frac{1}{\fpi}}
\def\fpi{f_\pi}
\def\mpi{m_\pi}
\def\mk{m_K}
\def\D{\Delta}
\def\AK{\overline{K}}
\def\pha{\Phi_a}
\def\phb{\Phi_b}
\def\phc{\Phi_c}
\def\phd{\Phi_d}
\def\ph0{\Phi_0}
\def\phs#1{\phi_S^{#1}}
\def\pia{\pi_a}
\def\pib{\pi_b}
\def\pis#1{\pi_S^{#1}}
\def\Ka{K_{\alpha}}
\def\Kb{K_{\beta}}
\def\AKa{K\dag_{\alpha}}
\def\AKb{K\dag_{\beta}}
\def\BKa{\AK_{\alpha}}
\def\BKb{\AK_{\beta}}
\def\Pkb{\Pi_\beta}
\def\JKa{J^K_\alpha}
\def\JKb{J^K_\beta}
\def\Jpa{J^\pi_a}
\def\Jpb{J^\pi_b}
\def\bK#1{{\cal K}_{#1}}
\def\JK#1{{\cal J}^K_{#1}}
\def\Jp#1{{\cal J}^\pi_{#1}}
\def\vA{{\bf A}}
\def\vI{{\bf I}}
\def\vL{{\bf L}}
\def\vO{{\bf O}}
\def\vP{{\bf P}}
\def\vQ{{\bf Q}}
\def\vR{{\bf R}}
\def\vS{{\bf S}}
\def\vT{{\bf T}}
\def\vX{{\bf X}}
\def\vY{{\bf Y}}
\def\vk{{\bf k}}
\def\vp{{\bf p}}
\def\vq{{\bf q}}
\def\vr{{\bf r}}
\def\vx{{\bf x}}
\def\vy{{\bf y}}
\def\vz{{\bf z}}
\def\vsig{\mbox{\boldmath{$\sigma$}}}
\def\vtau{\mbox{\boldmath{$\tau$}}}
\def\vOmega{{\bf \Omega}}
\def\vxi{\mbox{\boldmath{$\xi$}}}
\def\veta{\mbox{\boldmath{$\eta$}}}
\def\in{{\rm in}}
\def\out{{\rm out}}
\def\der{\partial}
\def\dag{^\dagger}
\def\eps{\varepsilon}
\def\half{\frac{1}{2}}
\def\del{\delta}
\def\ab{{\alpha\beta}}
\def\ba{{\beta\alpha}}
\def\G#1{G_{#1}}
\def\Ginv#1{{G^{-1}}_{#1}}
\def\X#1{X_{#1}}
\def\omq{\omega_q}
\def\omk{\omega_k}
\def\omkp{\omega_{k'}}
%

%%%%%%%%%%%%%%%%%%%%%%%%%%%%%%%%%%%%%%%%%%%%%%%%%%%%%%%%%%%%%%%%%%%
%section 1 
\section{Introduction}
\setcounter{equation}{0}
Two methods have been proposed in order to extend the $SU(2)$ 
Skyrme-soliton model to strange baryons, starting with 
the $SU(3)_L\times SU(3)_R$ chiral symmetric meson theory. 
One is the $SU(3)$ collective-coordinate method, which 
is a natural extension of the $SU(2)$ 
model\Ref{\ref{SU3model},\ref{Yabu}}.  
The other is the bound-state approach to strangeness, in which  the 
flavor $SU(3)$ symmetry in the baryon sector is not 
presumed\Ref{\ref{CK}}. \par

The nonlinear sigma model supplemented by the Wess-Zumino term is 
supposed to be an effective theory of QCD at the large $N_c$ limit, 
where $N_c$ is the number of colors\Ref{\ref{Witten}}.  Baryons are 
described in terms of solitons which are given as a classical solution 
of the nonlinear sigma model with appropriate stabilizing terms at 
leading order in the $1/N_c$ expansion. The Skyrme model has been 
revived as such a theory\Ref{\ref{AdNaWit}, \ref{AdkNapp}}. 
Thus, extension of the \SU{2} Skyrme-soliton model to 
the strange baryons should be consistent with the $1/N_c$ expansion in  
the large $N_c$ world. 
In this respect the bound-state approach to strangeness seems to be
more suitable for the extension of the \SU{2} soliton model to 
strangeness and other heavy flavors\Ref{\ref{Kaplan},\ref{DJM}}. \par

Since the proposal by Callan and Klebanov\Ref{\ref{CK}} it has been 
found that the bound-state approach to strangeness works 
fairly well in describing the mass spectrum and magnetic moments of 
hyperons\Ref{\ref{CHK}$\sim$\ref{Boffi}}. Kaon-nucleon background 
scattering has also been studied\Ref{\ref{Scoccola}}. There have been 
attempts to obtain the kaon coupling constants in the bound-state 
approach\Ref{\ref{Gobbi},\ref{Schat}}. However, meson-baryon scattering 
amplitudes including the strangeness exchange processes have 
not fully been developed so far within this approach. \par

In this paper we calculate the kaon coupling constants at 
hyperon-nucleon vertices and the pion ones at hyperon-hyperon 
vertices, where the positive-parity hyperons such as $\Lambda$, 
$\Sigma$ and $\Sigma(1385)$ denoted as $\Sigma^*$ are the P-wave bound 
states of antikaon to  the \SU{2} soliton and $\Lambda(1405)$ denoted as 
$\Lambda^*$ is the S-wave bound state. We also discuss the possible 
existence of negative-parity $\Sigma$ states with spin 1/2 and 3/2, 
which are induced from the S-wave bound kaon. The kaon and pion coupling 
constants are properly defined as matrix elements of the source terms 
of the mesons sandwiched between the two single-baryon states according 
to the prescription developed for resolving the Yukawa coupling 
problem in the \SU{2} Skyrme-soliton model\Ref{\ref{HSUnew},\ref{SU}}. 
Our method is simple, transparent and applicable to a wide range of 
meson vertices. Since we are interested in formulating the meson-baryon 
vertices including heavy quantum numbers, we restrict 
ourselves to the simplest Lagrangian which preserves the essentials of 
the bound-state approach.  
\par

In order to make numerical calculations we take the two 
sets of the parameters of the model, the pion decay constant $\fpi$ and 
the Skyrme constant $e$; Set I consists of $\fpi=54$MeV and $e=4.84$, 
and Set II does of $\fpi=93$MeV and $e=4.0$.  The parameters in Set I 
have been tuned so as to fit the masses of the nucleon and the 
$\Delta$ isobar by Adkins et al.\Ref{\ref{AdkNapp}}. In Set II the pion 
decay constant is kept equal to the physical value and the Skyrme 
constant is taken so as to give a reasonable size of the soliton and 
to reproduce the mass difference between the nucleon and the $\Delta$ 
isobar. Although the soliton has a large classical mass in Set II, 
it is reduced to a reasonable value if the one-loop corrections of 
\Nc{0} are taken into account\Ref{\ref{Moussallam},\ref{Holtzwarth}}; 
for example, the resultant Skyrmion mass is 873 MeV according to 
ref.\Ref{\ref{Holtzwarth}}. 
\par

The sizes of the pseudovector coupling constants $f_{YNK}$ to the 
positive hyperons are found to be close to those given by the 
compilation of the coupling 
constants\Ref{\ref{Edition},\ref{Timmermann}}.  The coupling 
constant $G_{\Lambda^*NK}$ in Set II is  close to  the phenomenological 
value\Ref{\ref{CHLee}}, but the one in Set I may be too large.  
Our results suggest that the parameter set II is 
more favorable than those of Set I. \par

In the next section the Lagrangian and Hamiltonian in the bound-state 
approach are given. The matrix element of the Hamiltonian in the 
intrinsic frame of the soliton is discussed in Appendix. The kaon and 
pion coupling constants to the positive parity hyperons are defined in 
the section 3. The kaon coupling to $\Lambda^*$ is given, and the 
possibility of $\Sigma$ states with negative parity predicted by the 
model are discussed in section 4. The conclusions and discussion are 
given in the last section. 
%%%%%%%%%%%%%%%%%%%%%%%%%%%%%%%%%%%%%%%%%%%%%%%%%%%%%%%%%%%%%%%%%%%
%section 2
\section{Lagrangian and Hamiltonian}
\setcounter{equation}{0}
For the sake of being self-contained, we give the Lagrangian and 
the Hamiltonian in this section. We start with the chiral 
$SU(3)\times SU(3)$ symmetric Lagrangian broken only by finite masses of 
the pion and kaon. The kaon fields are introduced as the fluctuations 
around the \SU{2} soliton according to the standard Callan-Klebanov 
ansatz\Ref{\ref{CK}}: 
\beq
U=\sqrt{U_\pi}\,U_K\sqrt{U_\pi}. \label{CKansatz}
\eeq
The kaon  part is written as 
\beq
U_K=\exp\left\{i\frac{\sqrt{2}}{\fpi}\left(
\begin{array}{lr}
0& K\\
K\dag& 0
\end{array}
\right)\right\},
\eeq
where 
\beq
K=\left(
\begin{array}{c}
K^+\\
K^0
\end{array}
\right)\qquad K\dag=(K^-,{\AK}^0), 
\eeq
and the pion part is 
\beq
U_\pi=\left(\begin{array}{rl}
u& 0\\
0& 1
\end{array}
\right)
\eeq
with 
\beq
u=\frac{1}{\fpi}(\ph0+i\tau_a\pha),\label{total}
\eeq
where $\pha$ is the total pion field, which consists of the classical 
Skyrmion configuration of \Nc{1/2} and the fluctuation field of \Nc{0} 
under the constraint $\ph0^2=\fpi^2-\sum\pha^2$\Ref{\ref{HSUnew}}. 
\par
The Lagrangian density $\CL$ is written as 
\beq
\CL=\CL_{sky}+\CL_K+{\rm O}(K^3),
\eeq
where $\CL_{sky}$ is the \SU{2} Skyrme Lagrangian, 
\beq
\CL_{sky}=\half\dot\pha \G{ab}\dot\phb-\CV[\pha,\nabla\pha],
\eeq
where the explicit expressions of $\G{ab}$ and $\CV$ are not given here. 
$\CL_K$ is bilinear in $K\dag$ and $K$: 
\beqa
\CL_K&=&(D_\mu K)\dag(D^\mu K)-\mk^2K\dag K
+\half\mpi^2(1-\frac{\ph0}{\fpi})K\dag K
-\frac{1}{4\fpi^2}K\dag K\left\{\frac{}{}\der_\mu\pha X_{ab}\der^\mu\phb 
\right.\noeq\\
&&\left.-\frac{2}{\fpi^2\kappa^2}\left[(\der_\mu\pha X_{ab}\der^\mu\phb)
(\der_\nu\phc X_{cd}\der^\nu\phd)-(\der_\mu\pha X_{ac}\der_\nu\phc)
(\der^\mu\phb X_{bd}\der^\nu\phd)\right]\right\}\noeq\\
&&+\frac{1}{4\kappa^2\fpi^2}\left\{
(D_\mu K)\dag(D_\nu K)(\der^\mu\pha X_{ab}\der^\nu\phb)
-(D_\mu K)\dag(D^\mu K)(\der_\nu\pha X_{ab}\der^\nu\phb)\right.\noeq\\
&&\left.-(D_\mu K)\dag\left[3i\eps_{abc}\der^\mu\pha\der^\nu\phb\tau_c
+6\fpi(V^\mu\der^\nu\ph0-V^\nu\der^\mu\ph0)\right](D_\nu K)\right\}
\noeq\\
&&-i\frac{N_c}{4\fpi^2}B^\mu\left\{K\dag(D_\mu K)-(D_\mu K)\dag K
\right\},
\eeqa
where $\kappa=e\fpi$ and 
\beqa
D_\mu&=&\der_\mu+V_\mu,\\
V^\mu&=&\frac{1}{2\fpi^2}\frac{1}{1+\ph0/\fpi}
i\eps_{abc}\tau_a\phb\der^\mu\phc,\\
B^\mu&=&\eps^{\mu\nu\alpha\beta}\frac{1}{24\pi^2}
\Tr(u\dag\der_\nu uu\dag\der_\alpha uu\dag\der_\beta u)\quad \mbox{with}
\quad \eps^{0123}=-1,\\
X_{ab}&=&\del_{ab}+\frac{\pha\phb}{\ph0^2}.
\eeqa
\par
In order to get the Hamiltonian we extract the time-derivative terms 
from $\CL_K$; 
\beqa
\CL_K&=&\dot K\dag f\,\dot K+\dot K\dag\eta_a\dot\pha+\dot\pha\eta_a\dag
\dot K\noeq\\
&&+\dot K\dag i\lambda\,K-K\dag i\lambda\dot K-i\Xi_a\dot\pha\noeq\\
&&+\dot\pha\widetilde{G_{ab}}\dot\phb+\CL',
\eeqa
where $\CL'$ does not include any time-derivatives of the kaon and pion 
fields, and 
\beqa
f&=&1+\frac{1}{4\kappa^2\fpi^2}\left(\der_i\pha X_{ab}\der_i\phb\right),
\label{f}\\
\lambda&=&\frac{N_c}{4\fpi^2}B^0,\label{lambda}\\
\eta_a&=&f\,V^0_aK+\frac{1}{4\kappa^2\fpi^2}\left\{\frac{}{}
-X_{ab}\der_i\phb+3i\eps_{abc}\der_i\phb\tau_c\right.\noeq\\
&&\left.+6\fpi\left(V^0_a\der_i\ph0+V_i\frac{\pha}{\ph0}\right)\right\}
(D_iK),\label{eta}\\
\Xi_a&=&2K\dag\left(\lambda V^0_a-\lambda_i^aV_i\right)K
+\lambda^a_i\left(\der_i K\dag\,K-K\dag\der_iK\right). \label{Xi}
\eeqa
When $\pha$ is set to the leading classical configuration, the term 
$\widetilde{G_{ab}}$ bilinear in the kaon fields is of \Nc{-1}, but 
$\eta_a$ and $\Xi_a$ are of \Nc{-1/2}. We neglect the term 
$\dot\pha\widetilde{G_{ab}}\dot\phb$ hereafter. 
In the above  $V^0_a$ and $\lambda^a_i$ are defined as  
\beqa
V^0_a\dot\pha&=&V^0,\\
\lambda^a_i\dot\pha&=&\frac{N_c}{4\fpi^2}B_i.
\eeqa
\par
We  define the momentum fields canonically conjugate to  $K(x)\dag$ and 
$\pha(x)$ as
\beqa
\Pi(x)&=&\frac{\del\CL}{\del\dot K\dag(x)}=f\dot K+i\lambda K
+\eta_a\dot\pha,\\
\pia(x)&=&\frac{\del\CL}{\del\dot\pha(x)}=\G{ab}\dot\phb-i\Xi_a
+\dot K\dag\eta_a+\eta_a\dag\dot K.
\eeqa
Note that $\dot\pha$, $\dot K$ and $\dot K\dag$ are solved 
in terms of $\pi$, $\Pi$ and $\Pi\dag$ without any constraints; 
\beqa
\dot K&=&f^{-1}[\Pi-i\lambda K-\eta_a\Ginv{ab}\pib]+{\rm O}(K^3),
\label{dotK}
\\
\dot\pha&=&\Ginv{ab}\left\{\pib+i\Xi_b 
-\left(\eta_b\dag f^{-1}[\Pi-i\lambda K-\eta_c\Ginv{cd}\pi_d]
+h.c.\right)\right\}.\label{dotpha}
\eeqa
Then the canonical commutation relations hold among the fields;
\beq
[\pha(\vx,t),\pib(\vy,t)]=i\del_{ab}\del(\vx-\vy),
\eeq
\beq
[\AKa(\vx,t),\Pkb(\vy,t)]=i\del_{\alpha\beta}\del(\vx-\vy).
\eeq
This is because the massive kaon fields do not contain the zero-mode 
wave functions of the \SU{2} Skyrmion configuration, and because the 
pion fields are the total fields\Ref{\ref{HSUnew}}.\par

We can construct the Hamiltonian density through the conventional 
method as 
\beq
\CH=\CH_{sky}+\CH_K+\CH_{\pi K},
\eeq
where $\CH_{sky}$ is  the one of the \SU{2} Skyrme model 
\beq
\CH_{sky}=\half(\pia\Ginv{ab}\pib)+\CV[\phi,\nabla\phi].\\
\eeq
The Hamiltonian density $\CH_K$ is given as 
\beqa
\CH_K&=&[\Pi\dag+iK\dag\lambda]f^{-1}[\Pi-i\lambda K]
+(D_iK)\dag d_{ij}(D_j K)\noeq\\
&&+[\mk^2-\half\mpi^2(1-\frac{\ph0}{\fpi})-v_0]K\dag\,K,
\eeqa
with 
\beqa
d_{ij}&=&\del_{ij}f+ \frac{1}{4\kappa^2\fpi^2}\left\{
-\der_i\pha X_{ab}\der_j\phb\right.\noeq\\
&&\left.+3i\eps_{abc}\der_i\pha\der_j\phb\tau_c+
6\fpi(V_i\der_j\ph0-V_j\der_i\ph0)\right\},\\
v_0&=&\frac{1}{4\fpi^2}\left\{\der_i\pha X_{ab}\der_i\phb+
\frac{2}{\kappa^2\fpi^2}[(\der_i\pha X_{ab}\der_i\phb)
(\der_j\phc X_{cd}\der_j\phd)\right.
\noeq\\
&&\left.-(\der_i\pha X_{ac}\der_j\phc)
(\der_i\phb X_{bd}\der_j\phd)]\frac{}{}\right\}.
\eeqa
The last part, $\CH_{\pi K}$, is linear in $\pia$ and written as 
\beq
\CH_{\pi K}=i\pia\Ginv{ab}\Xi_b
-\left\{\left(\pia\Ginv{ab}\eta_b\right)\dag 
D^{-1}\left(\Pi-i\lambda K\right)+\mbox{h.c.}\right\}.
\eeq
We should note that all the fields are defined in the laboratory system. 
\par
The Hamiltonian written in terms of the fields in the intrinsic frame of 
the Skyrme-soliton is given in  Appendix. The bound state parameters 
are tabulated in Table I. The masses of the positive parity hyperons are 
less than the empirical values, and the mass difference between 
$\Lambda$ and $\Sigma$ is a little bit large for the both sets of the 
model parameters. On the other hand the mass difference between 
$\Lambda^*$ and  $\Lambda$ is a little bit small. But our aim is not 
to search for the best parameters fitting to the baryon masses in 
this paper. 
\par
\begin{center}
Table I
\end{center}
%%%%%%%%%%%%%%%%%%%%%%%%%%%%%%%%%%%%%%%%%%%%%%%%%%%%%%%%%%%%
%section 3 Kaon pion coupling constants
\section{Kaon and pion couplings to positive-parity hyperons}
\setcounter{equation}{0}
In order to get the scattering amplitudes for the kaon, we introduce 
the asymptotic fields of the kaon, $K_{\in}(x)$ and $K_{\out}(x)$, 
which satisfy the free field equations: 
\beqa
{\Ka}_{\,\in}(x)&=&\sum_\vk\frac{1}{(2\pi)^{3/2}\sqrt{2\omk}}
\left\{b_\alpha(\vk)e^{-ikx}+a\dag_\alpha(\vk)e^{ikx}\right\},\\
{K\dag_\alpha}_{\,\in}(x)&=&\sum_\vk\frac{1}{(2\pi)^{3/2}\sqrt{2\omk}}
\left\{a_\alpha(\vk)e^{-ikx}+b\dag_\alpha(\vk)e^{ikx}\right\}
\eeqa
with $\omk=\sqrt{\vk^2+\mk^2}$, and $a_\alpha(\vk)\,(b_\alpha(\vk))$ 
is the annihilation operator of the antikaon\,(kaon) of the in-state 
with isospin index $\alpha=1/2,\,-1/2$. The same forms  are 
defined for out-fields. The field $\Ka(x)$ in the previous section 
is the interpolating field from the in-state to the out-state. 
The similar in- and out-fields are introduced to the pion fields, where 
the total field $\pha(x)$ itself plays a 
role of the interpolating field from the in-state to the 
out-state\Ref{\ref{HSUnew}}.  
The single-baryon state with the definite spin, isospin and momentum is 
given as the rotating and translating  Skyrme-soliton including a 
bound-state antikaon, if the baryon has strangeness. The Fock space is 
spanned by the in- and out-states composed of the in- and out- creation 
operators of the mesons acting on the single-baryon state. The 
baryon state is not the eigenstate of $H=\int d^3x\CH$, but 
$<B(\vp')|H|B(\vp)>=E_B(\vp)\del(\vp'-\vp)$ as shown in the Appendix, 
where $E_B=M_B+\vp^2/2M_B$ and terms of \Nc{-2} are discarded. 
\par

Thus, the scattering amplitude can be written through the LSZ reduction 
formula\Ref{\ref{LSZ}} as follows;
\beqa
T_{\bar KN\to\bar KN}&=&i(2\pi)^3\int d^4x e^{ik'x}
<N((\vp')|T\left({\JKb}\dag(x)\JKa(0)\right)\noeq\\
&&+\del(x^0)[\dot\AKb(x),\JKa(0)]-i\omkp\del(x^0)[\AKb(x),\JKa(0)]
|N(\vp)>
\eeqa
for $\BKa(\vk)+N(\vp)\to\BKb(\vk')+N(\vp')$. The second line of the 
above expression consisting of the equal-time commutators is called 
the seagull term or the contact term. The factor $(2\pi)^3$ comes from 
the normalization of the baryon wave function.  Strangeness exchange 
scattering is described in terms of the kaon and pion source terms as 
follows;
\beqa
T_{\bar KN\to\pi Y}&=&i(2\pi)^3\int d^4x e^{iqx}
<Y((\vp')|T\left(\Jpb(x)\JKa(0)\right)\noeq\\
&&+\del(x^0)[\dot\phb(x),\JKa(0)]-i\omq\del(x^0)[\phb(y),\JKa(0)]
|N(\vp)>
\eeqa
for $\BKa(\vk)+N(\vp)\to \pi_b(\vq)+Y(\vp')$. In the above the kaon and 
pion source terms are defined as   
\beqa
\JKa(x)&=&\ddot\Ka+(-\nabla^2+\mk^2)\Ka(x),\\
\Jpa(x)&=&\ddot\pha+(-\nabla^2+\mpi^2)\pha(x).
\eeqa
\par
Inserting the single-baryon states into the time-ordered product term, 
we get the Born terms, the residues of which are written in terms of 
the kaon and pion source terms sandwiched between two 
single-baryon states. 
\par

\subsection{Kaon couplings}
The source term in this section is restricted to leading order in 
the $1/N_c$ expansion when it is sandwiched between the single-baryon 
states. The second derivative with respect to $t$ is written as  
\beq
\ddot K=f^{-1}\left\{-2i\lambda\dot K+\left(v_0
+\half\mpi^2(1-\frac{\ph0}{\fpi})-\mk^2\right) K+D_i(d_{ij}D_j K)
\right\} \label{ddotK}
\eeq
at \Nc{0} through the commutator $i[H_K,\dot K]$, where 
$H_K=\int d^3x\CH_K$, and 
\beq
\dot K=f^{-1}(\Pi-i\lambda K).\label{newdotK}
\eeq
The commutator with $H_{sky}$ gives higher order terms. Note that 
Eq.(\ref{ddotK}) is the equation of motion to $\Ka$ in the laboratory 
system. 
\par

When we  calculate the matrix element of the source term  
sandwiched between the hyperon and nucleon, the kaon fields are 
transformed into the fields defined in the intrinsic frame, while the 
pion fields are reduced to the classical Skyrmion fields in the 
tree approximation as follows:  
\beqa
K_\alpha(x)&=&A_{\alpha i}\bK{i}(\vx-\vX(t),t), \label{IntrK}\\
\pha(x)&=&R_{ai}(t)\hat\phi_i(\vx-\vX(t)),\\
\ph0(x)&=&\hat\phi_0(\vx-\vX(t)),\\
\hat\phi_0(\vr)&=&\fpi\cos F(r)\qquad\mbox{and}\qquad
\hat\phi_i(\vr)=\fpi\hat r_i\sin F(r),
\eeqa
where  $\hat\vr=\vr/r$ and $F(r)$ is the profile function of the 
Skyrmion, $\vA(t)$ the \SU{2} matrix as the collective coordinates for 
the iso-rotation, $R_{ai}$ the orthogonal rotation matrix, and $\vX(t)$ 
the center of the Skyrmion as another set 
of the collective coordinates for the translational motion. 
\par

The kaon field in the intrinsic frame is expanded as 
\beq
\bK{}(\vr,t)=\sum_N \left\{
b_N\,k_N(\vr)e^{-i\widetilde{\omega}_Nt}
+a\dag_N\,k_N^c(\vr)e^{i\omega_Nt}
\right\},\label{BoundK}
\eeq
where $N=\{\ell, T,T_3\}$ with $\ell$ being the orbital angular momentum 
and $T$ and $T_3$ being the quantum numbers of 
$\vT=\vL+\vtau/2$, and $a_N$ and $\omega_N\,(b_N$ and 
$\widetilde{\omega}_N)$ are the destruction operator and energy of the 
strangeness $S=-1\,(+1)$ kaon with the quantum number $N$. We only take 
the part associated with $S=-1$ in the eigenmode expansion, hereafter. 
The charge-conjugate eigenmode is written as 
\beq
k_N^c(\vr)=k_{\ell T}^*(r)\vY^c_{T\ell T_3}(\theta,\phi)
\eeq
with
\beq
\vY^c_{T\ell T_3}=\left(
\begin{array}{c}
\dspl{<T,T_3|\ell,T_3+\half;\half,-\half>Y^*_{\ell,T_3+1/2}(\theta,\phi)}
\\[0.5cm]
\dspl{
-<T,T_3|\ell,T_3-\half;\half,\half>Y^*_{\ell,T_3-1/2}(\theta,\phi)
},
\end{array}\right) 
\eeq
where $Y_{\ell m}(\theta,\phi)$ is the usual spherical harmonics. 
\par
We now calculate the matrix elements of $\JKa(0)$, function of 
$\Phi(0)$ and $K(0)$, sandwiched 
between $<Y|$ and $|N>$: 
\beqa
&&<Y(\vp')|\JKa[\Phi(0),K(0)]|N(\vp)>
=<Y(\vp')|\vA_{\alpha i}\JK{i}[\hat\phi(-\vX(0)),\bK{}(-\vX(0))]|N(\vp)>
\noeq\\
&&=\frac{1}{(2\pi)^3}\int d^3r e^{i\vk\vr}
<Y|\vA_{\alpha i}\JK{i}[\hat\phi(\vr),\bK{}(\vr)]|N> 
\equiv\frac{1}{(2\pi)^3}<Y|\vA_{\alpha i}\widetilde{\JK{i}}(\vk)|N> 
\eeqa
where $\vk=\vp'-\vp$. In the above the eigenstate of $\vX(0)$, $|\vr>$, 
is introduced and used is 
$<\vr|N(\vp)>=\exp(i\vr\vp)/(2\pi)^{3/2}|N>$\Ref{\ref{HSUnew}}. 
We denoted the source term in the intrinsic frame as 
$\JK{i}(\vr,0)$, which is written as  
\beqa
\JK{}(\vr,0)&=&\ddot\bK{}(\vr,0)+(-\nabla^2+\mk^2)\bK{}(\vr,0)\noeq\\
&=&\sum_Na\dag_N(-\omega_N^2-\nabla^2+\mk^2)k^c_N(\vr)
\eeqa
with $\omega_N$ being the bound state energy, where 
we discarded $\dot\vA$ and $\ddot\vA$, because they are of higher order 
in the $1/N_c$ expansion, and used the equation of motion to 
$\bK{i}(\vr)$.  
\par

For the positive parity hyperons we take values 
$\ell=1$ and $T=1/2$, and then 
\beq
\bK{P}(\vr,t)
=k_1^*(r)e^{i\omega_1t}\left(
\begin{array}{l}
a\dag_{1/2}\sqrt{\frac{2}{3}}Y^*_{11}
+a\dag_{-1/2}\sqrt{\frac{1}{3}}Y^*_{10}\\[0.5cm]
a\dag_{1/2}\sqrt{\frac{1}{3}}Y^*_{10}+a\dag_{-1/2}\sqrt{\frac{2}{3}}
Y^*_{1-1}
\end{array}
\right)\equiv k_1(r)e^{i\omega_1t}\vOmega_1(a\dag;\theta,\phi),
\eeq
where $k_1(r)$ is the radial wave function, and $a\dag_t$ with 
$t=\pm 1/2$ denotes $a\dag_{1,1/2,t}$.  
Thus, we can write the Fourier transform of $\JK{i}$ 
as
\beqa
\widetilde{\JK{}}(\vk)&=&\int d^3r e^{i\vk\vr}\JK{}(\vr)\noeq\\
&=&(\omk^2-\omega_1^2)\int d^3rj_1(kr)k_1(r)
i\vOmega_1(a\dag;\hat\vk),
\eeqa
where $\omega_1$ is the P-wave bound-state energy and $\hat\vk=\vk/k$. 
\par

Now, we define here the nucleon state as\Ref{\ref{AdkNapp}} 
\beq
|N>=|i_3,j_3>=\sqrt{\frac{2}{8\pi^2}}(-1)^{1/2+i_3}
D^{1/2}_{-i_3,j_3}(\Theta)|0>,
\eeq
and the hyperon $\Lambda$, $\Sigma$ and $\Sigma^*$ states as
\beqa
|Y>&=&|I,I_3;J,J_3>\noeq\\
&=&\sum_t<J,J_3|I,J_3-t;\half,t>\sqrt{\frac{2I+1}{8\pi^2}}
(-1)^{I+I_3}D^I_{-I_3,J_3-t}(\Theta)a\dag_t|0>,
\eeqa
where $\Theta$ denotes the three Euler angles of the 
iso-rotation\Ref{\ref{Gobbi},\ref{Kondo}}.  We express the \SU{2} 
iso-rotation $\vA$ as $\vA_{\alpha i}= D^{1/2}_{\alpha,i}(\Theta)$. 
Thus, we have 
\beq
<Y|\vA_{\alpha i}\widetilde{\JK{i}}(\vk)|N>=
\Lambda_{YN}i(\vsig\cdot\vk)\widetilde{G_1}(k),\label{Kvertex}
\eeq
where the vertex function $\widetilde{G_1}(k)$ for the P-wave kaon is 
given as 
\beq
\widetilde{G_1}(k)=\sqrt{4\pi}\frac{(\omk^2-\omega_1^2)}{k}
\int drr^2j_1(kr)k_1(r),
\eeq
and $\vsig$ should be replaced by the transition spin matrix, $\vS$,  
from $J=1/2$ to $J=3/2$  for $Y=\Sigma^*$, which we define  as 
$(S_i)_{mn}=<3/2,m|1,i;1/2,n>$. The coefficients, $\Lambda_{YN}$'s, 
are given in Table II, where we note that the minus sign is 
multiplied to the vertices with $K^-$ meson, because the correct 
isospin multiplet of the atikaon is $(\AK^0,-K^-)$, while our 
antikaon multiplet is $(K^-,\AK^0)$. 
\par

Fixing the common mass scale at $\mk$ for the kaon coupling constants, 
we define the pseudovector coupling constant $f_{YNK}/\mk$ as 
\beq
\frac{f_{YNK}}{\mk}=\sqrt{4\pi}\Lambda_{YN}\lim_{\omk\to\omega_1}
\frac{\omk^2-\omega_1^2}{k}\int drr^2j_1(kr)k_1(r),
\eeq
because the Born term has the pole at $\omk=M_Y-M_N$, that is 
$\omk=\omega_1$ at leading order in the $1/N_c$ expansion. 
Using the asymptotic form of the normalized bound-state wave function, 
\beq
k_1(r)\sim \alpha_1\frac{1+\kappa_1 r}{r^2}e^{-\kappa_1r}
\eeq
with $\kappa_1=\sqrt{\mk^2-\omega_1^2}$, we have 
\beq
\frac{f_{YNK}}{\mk}=\sqrt{4\pi}\Lambda_{YN}\alpha_1.\label{pvkaon}
\eeq
Note that the dimension of $\alpha_1$ is linear in length from the 
normalization condition on $k_1$. The pseudoscalar coupling constant 
$G_{YNK}$ is given by 
\beq
G_{YNK}=\frac{M_N+M_Y}{\mk}f_{YNK}.
\eeq
\par

According to the compilation of coupling constants of 
1982-Edition\Ref{\ref{Edition}} and Ref.\Ref{\ref{Timmermann}} the 
empirical coupling constants are given, respectively, as 
\beq
\begin{array}{rclrcl}
\dspl{|\frac{f_{\Lambda pK^-}}{\sqrt{4\pi}}|}&=&0.89\pm 0.10&\qquad

\dspl{|\frac{f_{\Sigma^0 pK^-}}{\sqrt{4\pi}}|}&<&0.43\pm 0.07,\\
[0.5cm]
\dspl{|\frac{f_{\Lambda pK^-}}{\sqrt{4\pi}}|}&=&0.94\pm 0.03&\qquad
\dspl{|\frac{f_{\Sigma^0 pK^-}}{\sqrt{4\pi}}|}&=&0.25\pm 0.05,
\end{array}
\eeq
while our results on $|f_{\Lambda pK^-}/\sqrt{4\pi}|$ are $1.35$ for 
Set I and $0.92$ for Set II, and $|f_{\Sigma^0 p\bar K^-}/\sqrt{4\pi}|$ 
are $0.45$ and $0.31$ for Set I and II, respectively. The parameter set 
II seems to be better than Set I.  These results are very encouraging 
to the model. We note that if we use the $F/D$ ratio at $N_c=3$, we get 
the $F/D=(3\sqrt{3}-1)/(3\sqrt{3}+3)=0.507$ from 
$\Lambda_{\Lambda N}$ and $\Lambda_{\Sigma N}$, that is not far from 
$1/\sqrt{3}=0.577$, which will be given later from the pion couplings 
to the hyperons. The coefficients $\Lambda_{YN}$ and the pseudovector 
coupling constants $f_{YN\pi}/\sqrt{4\pi}$ are summarized in Table II. 
 \par
\begin{center}
\underline{Table II}
\end{center}
\par

\subsection{Pion couplings}
In order to derive the pion source term we have to calculate the 
second derivative of the total pion field with respect to time: 
$\ddot\pha$ is given through the commutator with $H_{sky}$ as 
\beq
\ddot\pha=-\Ginv{ab}\frac{\del\CV[\Phi,\nabla\Phi]}{\del\phb}+
\mbox{terms with $\pi^2$}.
\eeq
Note that if $\pha$'s are replaced by the classical fields of \Nc{1/2}, 
$\del\CV/\del\phb=0$ is the equation of motion to the 
classical soliton configuration, and that the terms with $\pi^2$ are 
discarded, since they are of \Nc{-3/2}.  Thus, the leading pion 
source term  $\Jpa$ comes from $(-\nabla^2+\mpi^2)\pha$, which 
gives the pion coupling constants of \Nc{1/2} to the positive parity 
hyperons  as shown for the nucleon and $\D$ 
couplings\Ref{\ref{HSUnew}}. For the positive parity hyperons 
the leading source term of the pion is given as 
\beq
\Jpa(0)=(-\nabla^2+\mpi^2)\pha(0),
\eeq 
and then  the pion coupling constant is written as 
\beq
<Y'|\widetilde{\Jpa}(\vq)|Y>=<Y'|R_{ai}\widetilde{\Jp{i}}(\vq)|Y>,
\eeq
where the rotational matrix 
$R_{ai}=1/2\Tr(\tau_a \vA\tau_i \vA\dag)$ is represented by 
$(-1)^aD^1_{-a,i}(\Theta)$ as the function of the Euler angles of the 
iso-rotation, that is consistent with 
$A_{\alpha i}=D^{1/2}_{\alpha i}(\Theta)$ used previously, and  
\beq
\widetilde{\Jp{i}}(\vq)=iq_i\frac{\omq^2}{q}\int d^3rj_1(qr)
\fpi\sin F(r).
\eeq
Then we have 
\beq
<Y'|\widetilde{\Jpa}(\vq)|Y>=\Lambda_{Y'Y}
(i\vsig\cdot\vq)\,\vI^{I'I}_a\widetilde{G_\pi}(q)
\eeq
for $Y'=\Sigma$ and $Y=\Sigma$ or $\Lambda$, and $(I_a^{I'I})_{I'_3I_3}
=<I'\,I'_3|1a;I\,I_3>$ is the transition isospin matrix from 
$I$ to $I'$. For $Y'=Y=\Sigma$ we define the isospin matrix for $I=1$ 
as $(I_a)^{I'_3I_3}=\sqrt{2}<1I'_3|1a;1I_3>$. The vertex function 
$\widetilde{G_\pi}(q)$ is given as 
\beq
\widetilde{G_\pi}(q)=4\pi\frac{\omq^2}{q}\fpi\int drr^2j_1(qr)\sin F(r).
\eeq
For $Y'=\Sigma^*$ and $Y=\Sigma,\,\Lambda$, 
\beq
<Y'|\widetilde{\Jpa}(\vq)|Y>=\Lambda_{\Sigma^*Y}
(i\vS\cdot\vq)\,\vI^{I'I}_a\widetilde{G_\pi}(q). 
\eeq
The residue of the $\Sigma^*$ resonance in the elastic 
$\pi\Lambda\to\Sigma^*\to \pi\Lambda$ process is written as
\beqa
&&(\vS\cdot\vq')\dag(\vS\cdot\vq)\Lambda^2_{\Sigma^*\Lambda}
\widetilde{G_\pi}^2(q)=P_3(\vq',\vq)\Lambda^2_{\Sigma^*\Lambda}
\frac{1}{3}\widetilde{G_\pi}^2(q),\label{Extra}\\
&&P_3(\vq',\vq)=3(\vq'\cdot\vq)-(\vsig\cdot\vq')(\vsig\cdot\vq),
\eeqa
where $P_3(\vq',\vq)$ is the projection operator of $J^P=3/2^+$. Note 
that an extra factor 1/3 appears in the residue. 
\par

Setting the mass scale to the pion mass for the pion coupling constants, 
we define the pseudovector pion coupling constant $f_{Y'Y\pi}$ as 
\beq
\frac{f_{Y'Y\pi}}{\mpi}=4\pi\fpi\Lambda_{Y'Y}\lim_{\omq\to 0}
\frac{\omq^2}{q}\int drr^2j_1(qr)\sin F(r),
\eeq
because the Born term has the pole at $\omq=M_{Y'}-M_Y$, that is zero at 
leading order. The argument similar to Eq.({\ref{pvkaon}) gives 
the pion coupling constant as 
\beq
\frac{f_{Y'Y\pi}}{\mpi}=4\pi\Lambda_{Y'Y}\fpi\alpha_\pi,\label{pvpion}
\eeq
where we used the form of $F(r)$ for large $r$, 
\beq
F(r)\sim \alpha_\pi\frac{1+\mpi r}{r^2}e^{-\mpi r}.
\eeq
\par
We find that the pion coupling constants of the hyperons are near 
the empirical values; $f_{\Sigma\Lambda\pi}/\sqrt{4\pi}=0.20\pm0.01$ and 
$f_{\Sigma\Sigma\pi}/\sqrt{4\pi}=0.21\pm0.02$\Ref{\ref{Edition}}.   
The pseudovector coupling constants and the coefficients are summarized 
in Table III, where we also give those of nonstrange nucleon and $\D$. 
Since the ratio $f_{\Sigma^+\Sigma^0\pi^+}/f_{\Sigma^+\Lambda\pi^+}$ is 
equal to one, the $F/D$ ratio defined at $N_c=3$ becomes $1/\sqrt{3}$. 
\begin{center}
\underline{Table III}
\end{center}
\par\noindent
The coefficients give the same ratios of the pion coupling constants 
as the \SU{3} symmetry at the large $N_c$ limit\Ref{\ref{DJM}}; 
for example 
\beqa
({\Sigma^*}^+\to\Sigma^0\pi^+)/({\Sigma^*}^+\to\Lambda\pi^+)&=&-1/2
\noeq\\
(\Delta^{++}\to p\pi^+)/({\Sigma^*}^+\to\Sigma^0\pi^+)&=&-\sqrt{6}.
\noeq
\eeqa
The $NN\pi$ coupling constant $f_{NN\pi}$ is a little bit small for 
Set II than the one for Set I and the empirical value. We 
think, however, that the value of $f_{NN\pi}$ is much improved in 
Set II, since $\fpi$ is kept equal to the physical value and the axial 
vector coupling constant $g_A$ becomes 1.03 in Set II compared to 
$\fpi=54$ MeV and $g_A=0.65$ in Set I. 
\par
%%%%%%%%%%%%%%%%%%%%%%%%%%%%%%%%%%%%%%%%%%%%%%%%%%%%%%%%%%%%%%%%%
%section 4 
\section{Kaon coupling to $\Lambda^*(1405)$ and negative-parity 
$\Sigma$ states}
\setcounter{equation}{0}
In the bound-state approach to strangeness $\Lambda^*$ is  
the bound state of the S-wave kaon.  The S-wave bound state disappears 
as the kaon mass becomes small below the physical one, for example 
the zero-energy bound state appears near $\mk\sim 300$MeV for the 
parameter set II, that is, the bound-state pole on the physical sheet 
moves to a resonance pole on the unphysical sheet of the $\AK-N$ 
scattering amplitude. 
\par

The kaon wave function in the intrinsic frame is given as 
\beq
\bK{S}(\vr,t)
=k^*_0(r)e^{i\omega_0 t}\left(
\begin{array}{l}
a\dag_{-1/2} Y^*_{00}\\[0.5cm]
-a\dag_{1/2} Y^*_{00}
\end{array}
\right)\equiv k\dag_0(r)e^{i\omega_0 t}\vOmega_0(a\dag;\theta,\phi),
\eeq
where $k_0$ is the radial wave function, and $a\dag_t=a\dag_{0,1/2,t}$. 
The $\Lambda^*$ state is expressed as 
\beqa
|\Lambda^*;j_3>&=&\sum_t\sqrt{\frac{1}{8\pi^2}}D^0_{0,J_3-t}(\Theta)
a\dag_t|0>
\noeq\\
&=&\sqrt{\frac{1}{8\pi^2}}D^0_{0,0}(\Theta)a\dag_{J_3}|0>.
\eeqa
Then, we have 
\beq
<\Lambda^*;J_3|A_{\alpha i}\widetilde{\JK{i}}(k)|N;j_3>
=\frac{1}{\sqrt{2}}\widetilde{G_0}(k)\del_{J_3j_3},
\eeq
where 
\beq
\widetilde{G_0}(k)=\sqrt{4\pi}(\omk^2-\omega_0^2)
\int drr^2j_0(kr)k_0(r)
\eeq
with $\omega_0$ being the S-wave bound-state energy. The pseudoscalar 
kaon coupling constant to $K\Lambda^*N$ is of \Nc{0} as the same 
as the kaon coupling constants for the positive parity hyperons and 
given as follows: Since the pole is at 
$\omk=M_{\Lambda^*}-M_N=\omega_0+{\rm O}(N_c^{-1})$, we have 
\beqa
G_{\Lambda^*NK}&=&\sqrt{4\pi}\Lambda_{\Lambda N}
\lim_{\omk\to\omega_0}(\omk^2-\omega_0^2)
\int drr^2j_0(kr)k_0(r)\noeq\\
&=&\sqrt{4\pi}\Lambda_{\Lambda N}\alpha_0\kappa_0, 
\eeqa
where $\kappa_0=\sqrt{\mk^2-\omega^2_0}$ and asymptotically 
\beq
k_0(r)\sim \frac{\alpha_0\kappa}{r}e^{-\kappa_0r}.
\eeq
Note that  $\alpha_0\kappa_0$ is dimensionless.\par

We get 
\beq
G_{\Lambda^*pK^-}/\sqrt{4\pi}=-1.45\quad{\rm and}\quad -0.72
\eeq
for Set I and II, respectively. According to the phenomenological 
analysis of the $\Lambda^*$ resonance contribution to $KN$ and $\AK N$ 
scattering length gives it to be 0.75 and 0.58 for 
$\bar g^2_{\Lambda_R}=0.25$ and $0.15$, respectively, where 
$|G_{\Lambda^*pK^-}|=\bar g_{\Lambda_R}\mk/\fpi$
\Ref{\ref{CHLee}}. So, the coupling constant in Set I seems to be too 
large. However, the experimental value of ${\rm Im}A_{K^-p}\sim 0.7$ fm 
would restrict the upper bound of $G^2_{\Lambda^*pK^-}/\sqrt{4\pi}$ 
to about 0.6 for $\Gamma^*=50$ MeV with $\Gamma^*$ being the decay width 
of $\Lambda^*$,  because the imaginary parts 
coming from various channels sum up due to unitarity of the elastic 
amplitude and the imaginary part of the resonance amplitude is written 
as 
\beq
i\frac{\Gamma^*/2}{(M_{\Lambda^*}-M_N-\mk)^2+(\Gamma^*/2)^2}
G^2_{\Lambda^*pK^-}.
\eeq
\par
\bigskip
Since the \SU{2} soliton has the rotational $I=J$ band with $I$ being 
an integral number 0, 1, $\cdots$ for $S=-1$ channel, 
the bound-state approach generates $\Sigma$ states with negative parity 
besides the $\Lambda^*$ state. We denote these  states as 
$\Sigma^-_{1/2}$ and $\Sigma^-_{3/2}$ with spin 1/2 and 3/2, 
respectively. Here we examine whether that $\Sigma^-_{S}$ can 
interact with non-exotic channels such as $\AK N$ and $\pi Y$ channels. 
\par

The $\Sigma^-_{1/2}$ state is written as 
\beq
|\Sigma^-_{1/2};I_3,J_3>=\sqrt{\frac{3}{8\pi^2}}\sum_{t=\pm 1/2}
(-1)^{1+I_3}<\half,J_3|1,J_3-t;\half,t>D^1_{-I_3,J_3-t}a\dag_t|0>.
\eeq
The $\Sigma^-_{1/2}N\AK$ coupling constant is given as 
\beq
<\Sigma^-_{1/2};J_3.I_3|A_{\alpha i}\widetilde{\JK{i}}(k)|N;j_3>
=\Lambda_{\Sigma^-_{1/2}N}\widetilde{G_0}(k)\del_{J_3j_3}, 
\eeq
where $\Lambda_{\Sigma^0_{1/2}p}=1/\sqrt{2}$ and 
$\Lambda_{\Sigma^+_{1/2}p}=-1$. $\Sigma^-_{1/2}$ would couple to 
$\Lambda\pi$ and $\Sigma\pi$ as like as $G_{\Lambda^*\Sigma\pi}$, 
which are of \Nc{-1/2}.  Thus, 
the $\Sigma^-_{1/2}$ state can strongly interact with  ${\AK}^0p$ and 
$\pi\Sigma$, which are the elastic channels.  
\par
As to $\Sigma^-_{3/2}$ the model predicts that it cannot interact with 
the elastic $\AK N$ channel, but can do with the $\AK\Delta$ at \Nc{0}:
\beq
<Z^-_{3/2};I_3,J_3|A_{\alpha i}\widetilde{\JK{i}}(k)|\Delta;i_3,j_3>=
\Lambda_{\Sigma\Delta}\widetilde{G_0}(k)\del_{J_3j_3}.
\eeq
The pion couplings among the negative-parity states $\Lambda^*$, 
$\Sigma^-_{1/2}$ and $\Sigma^-_{3/2}$ occur at leading order \Nc{1/2};
for example,
\beq
<\Sigma^-_{3/2};I_3,J_3|J^\pi_a(q)|\Lambda^*;j_3>=-\frac{1}{\sqrt{3}}
i(\vS\cdot \vq)\tilde G_\pi(q),
\eeq
but it is also impossible to couple to the $\pi Y$ channels in the 
model. Thus, the $\Sigma^-_{3/2}$ could not interact with the elastic 
channels. 
\par
Is there a candidate for the $\Sigma^-_{1/2,1/3}$ state ? If $c_0$ is 
larger than $c_1$, the mass difference between $\Sigma^-_{1/2}$ and 
$\Lambda^*$ cannot be larger than the mass difference between 
$\Sigma$ and $\Lambda$. Therefore, $\Sigma^-_{1/2}$  
could not be attributed to the established $\Sigma(1750)$, because the 
mass spacing from $\Lambda^*$ is too large. There is an indication of 
an enhancement  near 1480 MeV in the $\AK^0p$ mass spectrum, whose 
spin and parity are not known\Ref{\ref{Engelen},\ref{Pdata}}. 
$\Sigma^-_{1/2}$ may be attributed to this $\Sigma(1480)$, though it is 
not yet established, and if $\Sigma^-_{1/2}$ exists realy, it would 
lie above but not far from the $\AK N$ threshold. The resonance 
above the $\AK N$ threshold contributes a positive value 
to the $K^- n$ scattering length. The mass difference between 
$\Sigma^-_{3/2}$ and $\Lambda^*$ would be larger than 
$c_0(M_\Delta-M_N)$, but since it does not interact 
with the elastic channels, it may be difficult to observe the 
$\Sigma^-_{3/2}$ state. 
 \par
%%%%%%%%%%%%%%%%%%%%%%%%%%%%%%%%%%%%%%%%%%%%%%%%%%%%%%%%%%%%%%%%%%%% 
%section 5
\section{Conclusions and discussion}
\setcounter{equation}{0}
We have formulated the pion and kaon coupling constants to baryons with 
strangeness within the bound-state approach to strangeness in the Skyrme 
soliton model. The positive parity hyperon $\Lambda(1115)$, 
$\Sigma(1192)$ and $\Sigma^*(1385)$ appear as the bound states of the 
P-wave kaon to the \SU{2} soliton, whereas the $\Lambda^*(1405)$ does 
as the S-wave bound state in this approach.  \par

The pion fields used in the Lagrangian and Hamiltonian are defined as 
the total fields consisting of the classical Skyrmion fields and the 
fluctuation. The kaon fields are 
introduced as the fluctuation around the \SU{2} soliton in the 
laboratory system according to the Callan-Klebanov ansatz\Ref{\ref{CK}}. 
The meson-baryon vertices are defined as the source terms 
of the pion and kaon fields sandwiched between the single-baryon states. 
The sandwiched source term is rewritten in terms of the variables in 
the intrinsic frame of the soliton; the pion field becomes the Skyrmion 
field and the kaon field does the bound-state one. The kaon coupling to 
the positive-parity hyperon is of nonrelativistic pseudovector 
type. The coupling constant is defined as the residue at the 
pole of the Born term. According to this definition the magnitude of the 
kaon coupling constant is controlled by the asymptotic behavior of 
the normalized bound-state wave function as like as the pion coupling 
constant controlled by the asymptotic behavior of the chiral angle. 
Thus, if the equation to the bound state is the same, irrespectively to 
the ansatz by Callan-Klebanov or Blom et al, the resultant coupling 
constant deos not depends on the ansatz adopted. \par

The order of the coupling constants in the $1/N_c$ expansion is such 
that $f_{YNK}$ is of \Nc{0}, $f_{Y'Y\pi}$ of \Nc{1/2} and  
$G_{\Lambda^*NK}$ is of \Nc{0}. We found that the kaon vertex to the 
hyperon is of nonrelativistic pseudovector type as the same as the 
$\pi NN$ vertex. Since the $\Lambda^*$ state is the bound state of the 
S-wave kaon, strangeness cannot flows within the baryon line from 
$\Lambda^*$ to the positive-parity hyperon states. In order to obtain 
$G_{\Lambda^*\Sigma\pi}$, we have to construct higher 
order pion source terms of \Nc{-1/2},  which are bilinear in $\AK\,K$. 
This elaborate task to construct the higher order source term of the 
pion will appear elsewhere.  
\par 
Our meson-baryon vertices  are much more simple and transparent 
than those given in Refs.\Ref{\ref{Gobbi},\ref{Schat}} 
with respect to the definition of the vertices. If we adopt the \SU{3} 
collective-coordinate method instead of the bound-state approach, the 
Lagrangian for the kaon fields as the fluctuation around the nonstrange 
soliton is written as  bilinear forms of the kaon fields as the same as 
the bound-state approach.  But it cannot straightforward give the 
Yukawa vertices, because the hyperons appear as  
the \SU{3} rotating soliton and do not involve any kaons at tree level 
as in the \SU{2} case for the pion vertices. The Yukawa coupling of the 
kaon occurs as higher order term coming from the fact that the rotating 
solitons breaks the equations of motion. Our method 
can be applicable to the \SU{3} collective-coordinate method as in the 
\SU{2} case\Ref{\ref{HSUold}}, though it would be much involved. 
\par

{}From the comparison of the calculated kaon and pion coupling ggnstants 
with the phenomenological analyses\Ref{\ref{Edition},\ref{Timmermann},
\ref{CHLee}}, we 
found that the parameters of the model in Set II are better than those 
in Set I. We also note that the magnitude of $g_A$ of the nucleon 
is $g_A=1.03$ in Set II, but it is 0.65 in Set I.  Although the 
value of $\fpi$ is fixed to the empirical value 93 MeV, the Skyrme 
constant is set to a rather small value, $e=4.0$ in Set II. Such a small 
value of the Skyrme constant could not be supported by the chiral 
perturbation theory in the meson sector, but it seems that the Skyrme 
Lagrangian with the Set II parameters, having no chiral six order terms, 
is effectively equivalent to the one with a large value of the 
Skyrme constant supplemented by the standard six order term 
$-1/2e_6^2B^\mu B_\mu$ \Ref{\ref{Moussallam},\ref{Holtzwarth}}.
The sizes of the kaon coupling constants in Set I in our method are 
about twice as large as those of Refs.\Ref{\ref{Gobbi}, \ref{Schat}}. 
\par 

Since the \SU{2} soliton has the rotational $I=J$ band with $I$ being 
an integral number 0, 1, $\cdots$ for $S=-1$ channel, 
the bound-state approach  generates baryons with the isospin such as 
$I\ge 2$ for the P-wave bound state and $I\ge 1$ for the S-wave bound 
state. All of the $\Sigma$ states with  negative parity are not exotic, 
because the $\Sigma^-_{1/2}$ can interact with the $\bar K N$ and 
$\pi Y$ channels as discussed in the previous section. 
If the low-lying $\Sigma$ states with negative parity 
predicted by the model are not observed at all, the prediction of 
the $\Sigma$ states with negative parity may be a defect of the 
bound-state approach to Skyrme model. It is important, therefore, to 
reveal experimentally low energy resonances in the $\AK N$ and $\pi Y$ 
channels for the validity of the model. Nevertheless, we emphasize that 
the model can be applicable to low energy physics in great variety
as an effective theory. \par 
\bigskip
 
This work was partially supported by a Grant-in-Aid for Scientific 
Research of Japan Ministry of Education, Science and Culture 
(No. 06640405).
%%%%%%%%%%%%%%%%%%%%%%%%%%%%%%%%%%%%%%%%%%%%%%%%%%%%%%%%%%%%%%%%%%
%Appendix
\section*{Appendix \quad  Hamiltonian in the intrinsic frame} 
\renewcommand{\theequation}%
     {A.\arabic{equation}}
\setcounter{section}{0}
If we sandwich the total Hamiltonian $H=\int d^3x\CH$ between two 
single-hyperon states $|Y(\vp)>$,
$$
<Y(\vp')|H[\Phi,\pi;K,\Pi]|Y(\vp)>,
$$
the total field $\pha$ and $\pia$ are reduced to the classical Skyrmion 
configuration in the tree approximation\Ref{\ref{HSUnew}}, 
and also the kaon fields to the bound-state fields with the specific 
angular momentum, both of which are defined in the intrinsic frame of 
the Skyrme soliton.  These are defined in Eqs.(\ref{IntrK}) to 
(\ref{BoundK}). The time-derivatives of them are given as
\beqa
\dot K(x)&=&\vA(t)\dot\bK{}+\dot \vA\bK{}-\dot\vX\cdot\nabla\bK{},\\
\dot \pha(x)&=&\dot R_{ai}\hat\phi_i-R_{ai}\dot\vX\cdot\nabla\hat\phi_i
\eeqa
with
\beq
\begin{array}{rcl}
\dspl{\vA\dag\dot \vA}&=&\dspl{\frac{i\tau_a}{2}\dot \Theta_a},\\[0.5cm]
\dspl{R_{aj}\dot R_{ai}}&=&\dspl{\eps_{jib}\dot\Theta_b},
\end{array}
\eeq
where $\dot\Theta_a$ is the angular velocity around the a-th iso-spin 
axis and $\vX$ is the center of the Skyrmion. Since the mixing terms of 
the rotational and translational modes vanish or are of higher order, 
$\dot\Theta_a$ and $\dot X_i$ are 
given separately by their conjugate momenta, which are defined as 
\beqa
I_a&=&\frac{\der L}{\der\dot\Theta_a}=\Lambda_S\dot\Theta_a
-c_\ell T_a,\\
P_i&=&\frac{\der L}{\der\dot X_i}
=M_s\dot X_i+P^K_i, \label{totalP}
\eeqa
where $I_a$ is the angular momentum of the soliton in the intrinsic 
frame,  $\vT$ is the spin of the bound kaon, 
$(T_a)_{tt'}=a_t\dag(\tau_a/2)a_{t'}$, and the constant 
$c_\ell$ depends on the angular momentum of the bound kaon, 
which are given as
\beqa
c_1&=&1-\int drr^2k_1^2\omega_1\left\{\frac{4}{3}(1+c)(f
+\frac{s^2}{2\kappa^2r^2})
-\frac{1}{\kappa^2r^2}\frac{d\,}{dr}\left(r^2F's\right)\right\},\\
c_0&=&1-\int drr^2k_0^2\omega_0\left\{\frac{4}{3}(1-c)(f+
\frac{s^2}{2\kappa^2r^2})
+\frac{1}{\kappa^2r^2}\frac{d\,}{dr}\left(r^2F's\right)\right\}.
\eeqa
For the linear momentum $\vP^K$ denotes the kaon momentum defined as 
\beq
\vP^K=-\int d^3r\{\hat\Pi\dag\nabla\bK{}+h.c.\},
\eeq
where $\hat\Pi$ is the momentum field in the intrinsic frame, which 
can be defined as $f\dot\Kb{}+i\lambda\Kb{}$. Thus, we can regard 
$\vP_S=\vP-\vP^K$ as the baryon momentum. We note that there are 
additional higher order terms in Eq.(\ref{totalP}), which  vanish for 
the bound-state kaon, because the bound-state kaon has a definite 
angular momentum and parity. 
\par
Then we have the intrinsic Hamiltonian responsible to the bound-states 
in the tree approximation: 
\beq
H=M_S+\sum_{\ell=0,1}\frac{(\vI+c_\ell\vT)^2}{2\Lambda_S}
+\frac{\vP^2_S}{2M_S}+\sum_{\ell=0,1,\,t=\pm 1/2}
 \omega_\ell a_{\ell,t}\dag a_{\ell,t}. \label{Hint}
\eeq
We see that $\CH_{\pi K}$ is absorbed into the first term through 
the transformation from the laboratory to the intrinsic frame. 
Thus, we see that 
\beqa
&&<Y(\vp')|H[\Phi,\pi;K,\Pi]|Y(\vp)>=\left(M_s+\omega_\ell
+\frac{(\vI+c\vT)^2}{2\Lambda_S}+\frac{\vp^2}{2M_s}\right)
\del(\vp'-\vp)\noeq \\
&&=\left(M_Y+\frac{\vp^2}{2M_Y}\right)\del(\vp'-\vp)+{\rm O}(N_c^{-2}),
\eeqa
which is the nonrelativistic energy of the hyperon in the tree 
approximation. 
\par
%%%%%%%%%%%%%%%%%%%%%%%%%%%%%%%%%%%%%%%%%%%%%%%%%%%%%%%%%%%%%%
%References
\newpage
%\vspace{2cm}
\baselineskip 24pt
\begin{center}
{\bf References}
\end{center}
\def\labelenumi{[\theenumi]}
%%%%%%%%%%%%%%%%%%%%%%%%%%%%%%%%%%%%%%%%%%%%%%%%%%%%%%%%%%%%%%%%%
\def\Ref#1{[{#1}]}
\def\npb#1#2#3{{Nucl. Phys.\,}{\bf B{#1}}\,(#3), #2}
\def\npa#1#2#3{{Nucl. Phys.\,}{\bf A{#1}}\,(#3),#2}
\def\np#1#2#3{{Nucl. Phys.\,}{\bf{#1}}\,(#3),#2}
\def\plb#1#2#3{{Phys. Lett.\,}{\bf B{#1}}\,(#3),#2}
\def\prl#1#2#3{{Phys. Rev. Lett.\,}{\bf{#1}}\,(#3),#2}
\def\prd#1#2#3{{Phys. Rev.\,}{\bf D{#1}}\,(#3),#2}
\def\prc#1#2#3{{Phys. Rev.\,}{\bf C{#1}}\,(#3),#2}
\def\pr#1#2#3{{Phys. Rev.\,}{\bf{#1}}\,(#3),#2}
\def\ap#1#2#3{{Ann. Phys.\,}{\bf{#1}}\,(#3),#2}
\def\prep#1#2#3{{Phys. Reports\,}{\bf{#1}}\,(#3),#2}
\def\rmp#1#2#3{{Rev. Mod. Phys.\,}{\bf{#1}}\,(#3),#2}
\def\cmp#1#2#3{{Comm. Math. Phys.\,}{\bf{#1}}\,(#3),#2}
\def\ptp#1#2#3{{Prog. Theor. Phys.\,}{\bf{#1}}\,(#3),#2}
\def\ib#1#2#3{{\it ibid.\,}{\bf{#1}}\,(#3),#2}
\def\zsc#1#2#3{{Z. Phys. \,}{\bf C{#1}}\,(#3),#2}
\def\zsa#1#2#3{{Z. Phys. \,}{\bf A{#1}}\,(#3),#2}
\def\intj#1#2#3{{Int. J. Mod. Phys.\,}{\bf A{#1}}\,(#3),#2}
\def\etal{{\it et al.}}
%%%%%%%%%%%%%%%%%%%%%%%%%%%%%%%%%%%%%%%%%%%%%%%%%%%%%%%%%%%%%%
\begin{enumerate}
\item \label{SU3model} E. Guadanini,\npb{236}{35}{1984};\\
P.O. Mazur, M.A. Novak and M. Praszlowicz, \plb{147}{137}{1984};\\
M. Plaszalowicz, \plb{158}{214}{1985}.
\item \label{Yabu} H. Yabu and K. Ando, \npb{301}{601}{1988}.
\item \label{CK} C.G. Callan and I. Klebanov, 
\npb{262}{367}{1985};
\item \label{Witten} E. Witten, \npb{223}{422}{1983}; 
{\bf B223}, 825\,(1983). 
\item \label{AdNaWit} G.S. Adkins, C.R. Nappi and E. Witten, 
\npb{288}{552}{1983}. 
\item \label{AdkNapp} G.S. Adkins and C.R. Nappi, \npb{B233}{109}{1984}. 
\item \label{Kaplan} D.B. Kaplan and I. Klebanov, \npb{335}{45}{1990}.
\item \label{DJM} R. Dashen, E. Jenkins and A.V. Manohar, \prd{49}{4713}
{1994}.
\item \label{CHK} C.G. Callan, K. Hornbostel and I. Klebanov, 
\plb{202}{269}{1988}.
\item \label{mass} 
N.N. Scoccola, H. Nadeau, M.A. Nowak and M. Rho, \plb{201}{425}{1988};\\
J.P. Blaizot, M. Rho and N.N. Scoccola, \plb{209}{27}{1988}.
\item \label{Blom} U. Blom, K. Dannbom and D.O. Riska, 
\npa{491}{384}{1989}.
\item \label{Nyman} E.M. Nyman and D.O. Riska, \npb{325}{593}{1989}.
\item \label{vector}
N.N. Scoccola, D.-P. Min, H. Nadeau and M. Rho, \npa{505}{497}{1989}.
\item \label{Kunz} J. Kunz and P.J. Mulders, \prd{41}{1578}{1990}.
\item \label{magnet} 
Y. Oh, D.-P. Min, M. Rho and N.N. Scoccola, \npa{534}{493}{1991}.
\item \label{Boffi} C. Gobbi, S. Boffi and D.O. Riska, 
\npa{493}{633}{1992}.
\item \label{Scoccola} N.N. Scoccola, \plb{236}{245}{1990}.
\item \label{Gobbi} C. Gobbi, D.O. Riska and N.N. Scoccola, 
\npa{544}{671}{1992}.
\item \label{Schat} C.L. Schat, N.N. Scoccola and C. Gobbi, 
\npa{585}{627}{1995}.
\item \label{HSUnew} A. Hayashi, S. Saito and M. Uehara, 
\prd{46}{4856}{1992}.
\item \label{SU} S. Saito and M. Uehara, \prd{51}{6059}{1995}.
\item \label{Moussallam} B. Moussallam, \ap{225}{264}{1993}.
\item \label{Holtzwarth} G. Holtzwarth and H. Walliser, 
\npa{587}{721}{1995}.
\item \label{Edition} O. Dumbrajs, R. Koch, H. Pilkuhn, G.C. Oades, 
H. Behrens, J.J. de Swart and\\ P. Kroll, \npb{216}{277}{1983}. 
\item \label{Timmermann} R.G.E. Timmermann, Th.A. Rijken and J.J. 
de Swart,\\ \npa{585}{143c}{1995}.
\item \label{CHLee} C.-H. Lee, H. Jung, D.-P. Min, and M.Rho, 
\plb{326}{14}{1994};\\
C.-H Lee, G.E. Brown, D.-P Min and M. Rho, \npa{585}{401}{1995}.
\item \label{LSZ} H. Lehmann, K. Symanzik and W Zimmermann, 
Nuovo Cim. {\bf 1}\,(1955),205. 
\item \label{Kondo} Y. Kondo, S. Saito and T. Otofuji, 
\plb{236}{1}{1990}.
\item \label{Engelen} J.J. Engelen et al., \npb{133}{61}{1980}.
\item \label{Pdata} Particle Data Group, \prd{50}{1753}{1994}, 
and refences therein.
\item \label{HSUold} A. Hayashi, S. Saito and M. Uehara, 
\prd{43}{1520}{1991}.
\end{enumerate}
%%%%%%%%%%%%%%%%%%%%%%%%%%%%%%%%%%%%%%%%%%%%%%%%%%%%%%%%%%%%%%%
%Figure and Tables
\newpage
\begin{center}
{\bf Tables and Table Captions}\par
\end{center}\par

\begin{description}
\item{Table I} \quad Bound-state parameters.  The parameters of 
the model are taken to be $\fpi=54$ MeV and $e=4.84$ in Set I, and  
$\fpi=93$ MeV and $e=4.0$ in Set II. The pion and kaon masses are 
taken as 138MeV and  495 MeV, respectively,  for the both sets.
\item{Table II}\quad  $\Lambda_{YN}$ and the pseudovector coupling 
constants $f_{YNK}$. The pseudoscalar coupling constant is given 
by $G_{YNK}=(M_Y+M_N)/\mk\cdot f_{YNK}$.
The empirical values with $^{*)}$ are taken from ref. 24), and 
those with $^{**)}$ from ref. 25). 
\item{Table III}\quad $\Lambda_{Y'Y}$ and the pseudovector coupling 
constants. $f_{NN\pi}$ and $f_{\Delta N\pi}$ are also given for 
comparison. $G_{Y'Y\pi}=(M_Y+M_{Y'})/\mpi\cdot f_{YY'\pi}$. 
The empirical values with $^{*)}$ are taken from ref. 24), 
where the empirical value $f_{\Sigma^*\Lambda\pi}$ is multiplied by the 
extra factor $\sqrt{3}$. ( See Eq.(\ref{Extra}).)
\end{description}
\par

\bigskip
\begin{center}
\renewcommand{\arraystretch}{2}
\begin{tabular}{|c|c|c|c|c|c|}\hline
 & $\omega_1$(MeV) & $\omega_0$(MeV) & $c_1$ & $c_0$ 
 &$1/\Lambda_S$(MeV) \\ \hline 
Set I & 147 & 339 & 0.513 & 0.816 & 195 \\ \hline
Set II & 183 & 434 & 0.388 & 0.788 & 167 \\ \hline
\end{tabular}
\end{center}
\par 
\begin{center}
Table I
\end{center}
\bigskip
\begin{center}
\renewcommand{\arraystretch}{2}
\begin{tabular}{|c||c|c|c|c|}\hline
& $\Lambda_{YN}$ &
\multicolumn{3}{c|}{$|f_{YNK}/\sqrt{4\pi}|$}\\ \cline{3-5}
& & Set I & Set II & empirical \\ \hline
$\dspl{\Lambda_{\Lambda p}}=\dspl{-\Lambda_{\Lambda n}}$ 
&$\dspl{1/\sqrt{2}}$ & 1.35 & 0.92 & 
$\begin{array}{c} 0.89\pm0.10^{*)}\\ 0.94\pm0.03^{**)}\end{array}$ \\ \hline
$\dspl{\Lambda_{\Sigma^+ p}}=\dspl{\Lambda_{\Sigma^-n}}$
&$-\dspl{1/3}$ &0.64 & 0.43 & \\ \hline
$\dspl{\Lambda_{\Sigma^0p}}=\dspl{\Lambda_{\Sigma^0n}}$ 
&$\dspl{-1/3\sqrt{2}}$ &0.45 & 0.31 & 
$\begin{array}{c} <0.43\pm0.07^{*)}\\ 0.25\pm0.05^{**)}\end{array}$\\ \hline
$\dspl{\Lambda_{\Sigma^{*+}p}}=\dspl{\Lambda_{\Sigma^{*-}n}}$
&$-\dspl{2/\sqrt{3}}$ & 2.21& 1.50 & \\ \hline
$\dspl{\Lambda_{\Sigma^{*0}p}}=\dspl{\Lambda_{\Sigma^{*0}n}}$
&$-\dspl{\sqrt{2/3}}$ & 1.55 & 1.06 & \\ \hline
\end{tabular}
\end{center}
\par
\begin{center}
Table II
\end{center}
\bigskip
\begin{center}
\renewcommand{\arraystretch}{2}
\begin{tabular}{|c||c|c|c|c|}\hline
$Y'Y$ & $\Lambda_{Y'Y}$ &
\multicolumn{3}{c|}{$|f_{Y'Y\pi}/\sqrt{4\pi}|$}\\
\cline{3-5}
& & Set I& Set II & empirical \\ \hline 
$\dspl{{ \Sigma\Lambda}}$ & $\dspl{1/3}$ 
&0.25 & 0.22 & 0.20$\pm$0.01$^{*)}$ \\ \hline
$\dspl{{\Sigma\Sigma}}$ & $\dspl{1/3}$ 
& 0.25 & 0.22 & 0.21$\pm$ 0.02$^{*)}$\\ \hline
$\dspl{{\Sigma^*\Lambda}}$ & $\dspl{-1/\sqrt{3}}$ 
& 0.43 & 0.38 & 0.25$\pm$ 0.01$^{*)}$\\ \hline
$\dspl{{\Sigma^*\Sigma}}$ & $\dspl{1/\sqrt{12}}$ 
& 0.21 & 0.19 & \\ \hline
$\dspl{{NN}}$& $-1/3$ & $ 0.25$ &0.22 & 0.27 \\ \hline
$\dspl{{\Delta N}}$ & $\dspl{-1/\sqrt{2}}$ & 0.54 
&0.47 & 0.47 \\ \hline
\end{tabular}
\end{center}
\par
\begin{center}
Table III  
\end{center}
\par
%%%%%%%%%%%%%%%%%%%%%%%%%%%%%%%%%%%%%%%%%%%%%%%%%%%%%%%%%%%%%%%%%%
\end{document}